\documentclass[aps]{revtex4}
\usepackage{graphicx}
\usepackage{bm}
\usepackage{amsmath}
\usepackage[T1]{fontenc}
\usepackage{mathrsfs,ae,dsfont}

\begin{document}

\title{A quantum description for charged fermions in strong
gravitational field}

\author{Hai-Jhun Wanng}

\affiliation{Center for Theoretical Physics and School of Physics,
Jilin University, Changchun 130012, China}

\begin{abstract}
The falling charge puzzle in gravitational field is well known due
to the discussions of radiation. The puzzle lies in the heart of
linking the electromagnetism and gravity. Up to date few
discussions have fully taken account of quantum effect of a
falling charged-fermion in strong gravitational field from the
first principle. Based on the hypothesis that 4-dimension
conformal symmetry may underly its dynamics, in this paper we try
to establish a quantum equation for the falling process. The
resultant equation provides a manner accounting for the strong CP
violation at the beginning of the Big Bang. Moreover, it turns out
that the equation for left-handed fermions breaks the conformal
symmetry and has a tensor-like eigen value. A proposed experiment
for testing the predictions is also suggested.
\end{abstract}

\maketitle

\section{Introduction}


A charge is accelerated in gravitational field due to its mass,
and the accelerated charge occurs bremsstrahlung. So one infers
that in certain cases a charge propagating in strong gravitational
field radiates. In phrase "strong gravitational field" the
"strong" means the gravitational field strong enough to accelerate
the mass of the charge comparable to that of common electronic
field accelerating the charge. Abundant aspects of the "falling
charge" have been discussed ever since 1909 ~\cite{Born1909}. In
most previous studies people relied on the equivalence principle
(EP) of general relativity, which means a reference frame able to
co-move with the charge while it propagating in gravitational
filed~\cite {Obukhov11}. The famous puzzle then occurs that from
classical electrodynamics an accelerated charge radiates, whereas
a co-moving observer would observe no radiation according to EP,
i.e. an observer sitting in
different reference frames would observe conflicting phenomena, see ref.~%
\cite{Fulton1960, Rohrlich63, Fulton62}. By viewing the radiation
all be photons quantized one by one, Feynman proposed mentally a
machine detecting photons in different reference
frames~\cite{Feynman95}, the classically ''comoving'' observer
could also recognizes the radiation by reflection, which gives the
same number of radiating photons by counting. This
Gedanken-Experiment makes the puzzle more elusive.


~~\\

Previous investigations have fallen into three categories, the
first category focuses on the classical motion of a charge in
gravitational field~ \cite{Born1909, Dirac38}, the second focuses
on the self-interaction of the charge and how it affects the
radiation~\cite{DeWitt60, Rohr00, Rohr00A}. The third is named
quantum field theory in curved space-time by introducing the
Christoffel tensor $\Gamma _\mu $ into covariant differential operator $%
D_\mu =\partial _\mu -\Gamma _\mu $ ~\cite{Niko03, Niko05}. The first
focused once on the problem why a supported charge doesn't radiate~\cite
{Rohrlich63}, and how a uniformly accelerated charge corresponds to
gravitation~\cite{Fulton1960, Lyle08}. This uniformly accelerating analogy
just makes paedological sense since sooner or later the charge would have
velocity more than light~\cite{Papini15}. The second category has been
generalized by authors to a broad category consisting of any
self-interaction system, including gravitation ~\cite{Barack19} and
electromagnetic system ~\cite{Crem15} as well as others. There are other
attempts to compute the radiation of a charge in gravitational field (CGF)
by employing Dirac-Lorentz equation etc.~\cite{Rohr08}. In previous
literature quantum effect has been involved with different manners. So far a
thorough quantum-mechanical-description from the first principle is called
for to clarify the mist and to resolve the puzzle.


~~\\

Purely quantum mechanical interpretation of CGF may differ from
those of classical in that for a quantum particle the comoving
reference frame would not exist due to the Heisenberg's
uncertainty principle. This point was not pointed out by previous
investigations. A charged particle accelerated by gravitational
field, without comoving classical reference frame nor other
charges to bind it to form atoms like hydrogen, would inevitably
radiate. We also suppose there is no screening phenomenon at the
very short moment concerned. In such case we have to construct a
new equation of quantum motion (EoM) for the accelerated quantum
system in the place of Klein-Gordon equation or Dirac equation
since they lose their effectiveness without inertial reference
frame. To this end, it may be a good start to consider the
symmetric feature of the system. The gravitational field possesses
the largest symmetry of the 4-dimension space-time, i.e. conformal
symmetry. So we may follow the symmetry to construct the equation
for fermions, like the process of constructing Dirac equation,
except that now we work in the spinor representation of the
4-dimensional conformal group~\cite{Han15}, a larger group than
Lorentz group.

~~\\

The remainder of this paper are arranged as follows, in Sect. II
we devote to find an equation describing the EOM of CGM, finally
we find out an equation for a left-handed fermion falling in
strong gravitational field. The equation has tensor-like
eigen-value. In Sect. III we find out the operator in quantum
equation for fermions can provide a manner accounting for the
strong CP violation at the beginning of the Universe. In Sect. IV
we suggest a route how to construct an operator equation based on
the knowledge of previous sections. Finally we present the
conclusions and discussions.

~~\\

\section{The Conformal Methodology in Constructing New Quantum Equation}

For a classical neutral particle its trajectory in gravitational field is
called geodesic line, along which the particle itself feels no force.
Correspondingly a co-moving observer would observe no acceleration according
to EP, which is the intuitive interpretation of EP. For a charged particle
however, the geodesic line would not exist since the radiation accompanies
the acceleration. So the EoM of classical neutral particle does not pertain
to the charged particle any longer. Here we attempt to formulate a new
equation. More concretely, let's consider a unit charge taken by a fermion.
From quantum viewpoint, it is inappropriate to say the trajectories of a
fermion. Instead we prefer to describe it with wave function rather than
with its trajectory, thus we need an equation for the wave function. The
construction of the equation may mimick the process of constructing the
Dirac equation. However, the Dirac equation meets the condition of
energy-momentum conservation $E^2=m^2+\vec p^2$, whereas in the case of a
falling charge in strong gravitational field we note its energy-momentum
doesn't have such good relation due to the radiation. So the new equation
must exceed the scope of Dirac equation and the scope of energy-momentum
conservation as well. Although there is no geodesic line for charged
particles in gravitational field, the conformal symmetry should be followed
since it is the largest symmetry for 4-dimension dynamics.

~~\\

One popular stream combining gravity and electromagnetism was to take into
account covariant differential $\nabla _\mu =\partial _\mu -\Gamma _\mu $,
where $\Gamma _\mu $ is the Levi-Civita connection used in general
relativity (GR). Most of previous literatures extended $\nabla _\mu
=\partial _\mu -eA_\mu $ to $\nabla _\mu =\partial _\mu -A_\mu -\Gamma _\mu $
~\cite{Fulton62, Obukhov11}. This interpretation proves to be effective
while the gravitational field is very weak ~\cite{Colella75}. The expression
is also supposed to validate the quantization of gravity. Actually, this
expression turns out to find the quantum aspect of gravity, since in EoMs $%
\Gamma _\mu $ is also performed on quantum wave functions. Another stream,
if we view electromagnetic filed and gravitational field on an equal footing
in classical cases, as implied in $R^{\mu \nu }-\frac 12g^{\mu \nu }R=\sigma
T^{\mu \nu }$, where the tensor of electromagnetic filed is included in $%
T^{\mu \nu }$, then actually the electromagnetic potential $A_\mu $ behaves
as normal four-vector as any other classical vectors, focusing on the
gravitational effects caused by electromagnetic strength tensor. In quantum
theory one has employed the former and in classical GR one is used to the
latter. And in the former one draws $\Gamma _\mu $ to quantum level to see
its quantum effects but without the interaction between $\Gamma _\mu $ and $%
A_\mu $, and in the latter one draws $A_\mu $ to classical level, omitting
the induction aspects between $\vec E$ and $\vec B$, just viewing them as
normal energy-momentum tensor. We recognize these two methods have nothing
to do with our goal of discussing radiation in strong gravitational field,
also it is difficult to generalize them to meet conformal symmetry, thus we
will follow neither of them.

~~\\

A straightforward way of building an conformal-invariant equation is to
study prudently the coupling of two representations of conformal group, the
spinor representation $\{\Gamma _\lambda \}$ and its original differential
representation $\{\Lambda ^\lambda \}$, as $\sum_\lambda \{\Gamma _\lambda
\Lambda ^\lambda \}=$something or $\sum_{\sigma ,\rho }\{g_{\sigma \rho
}\Gamma ^\sigma \Lambda ^\rho \}=$something ($g_{\sigma \rho }$ marks the
metric that keeps the conformal invariance, $ds^2=g_{\sigma \rho }dx^\sigma
dx^\rho $), just like the way of finding Dirac equation $\gamma _\mu
\partial ^\mu \,\psi =m\,\psi $, where $\{\Gamma _\lambda ,\lambda
=1,2,\cdots \cdots ,15\}$ are the set of generalized gamma matrices, and the
differential operators $\{\Lambda ^\lambda \}$ are as follows
\begin{eqnarray}
D &=&i\,x_\mu \frac \partial {\partial x_\mu }\text{, }M_{\mu \nu }=i(x_\mu
\frac \partial {\partial x^\nu }-x_\nu \frac \partial {\partial x^\mu })%
\text{,}  \nonumber \\
P_\mu &=&i\frac \partial {\partial x^\mu }\text{ , }K_\mu =-i(x^2\frac
\partial {\partial x^\mu }-2x_\mu x^\nu \frac \partial {\partial x^\nu })%
\text{,}  \label{ccc}
\end{eqnarray}
where $M_{\mu \nu }$ represent the components of conventional angular
momentum in 4-dimension. The corresponding commutation relation of above
operators can be checked directly,
\begin{eqnarray}
\lbrack M_{\mu \nu },M_{\rho \sigma }] &=&i(\eta _{\nu \rho }M_{\mu \sigma
}+\eta _{\mu \sigma }M_{\nu \rho }-\eta _{\mu \rho }M_{\nu \sigma }-\eta
_{\nu \sigma }M_{\mu \rho }),  \nonumber \\
\lbrack M_{\mu \nu },P_\rho ] &=&i(\eta _{\nu \rho }P_\mu -\eta _{\mu \rho
}P_\nu ),  \nonumber \\
\lbrack D,P_\mu ] &=&-iP_\mu \text{, }[D,K_\mu ]=iK_\mu ,  \nonumber \\
\lbrack D,M_{\mu \nu }] &=&0  \nonumber \\
\ \ [M_{\mu \nu },K_\rho ] &=&i(\eta _{\nu \rho }K_\mu -\eta _{\mu \rho
}K_\nu )  \nonumber  \label{dd} \\
\ [P_\mu ,K_\rho ] &=&-2\,i\,(\eta _{\mu \rho }\,D+M_{\mu \rho })
\label{Commu}
\end{eqnarray}
The above tensors $\eta _{\mu \rho }$ are just those in special relativity,
\[
(\eta _{\mu \rho })=\left(
\begin{array}{cccc}
1 & 0 & 0 & 0 \\
0 & -1 & 0 & 0 \\
0 & 0 & -1 & 0 \\
0 & 0 & 0 & -1
\end{array}
\right) \text{ .}
\]
The corresponding operators $\{\Gamma _\lambda \}$ in spinor representations
are ~\cite{Han15} ,
\begin{eqnarray}
M_{jk} &\longrightarrow &i(x_j\frac \partial {\partial x^k}-x_k\frac
\partial {\partial x^j})\longrightarrow \gamma _i\gamma _j  \nonumber \\
M_{0k} &\longrightarrow &i(x_i\frac \partial {\partial x^0}-x_0\frac
\partial {\partial x^i})\longrightarrow \gamma _0\gamma _i  \nonumber \\
D &\longrightarrow &i\,x_\mu \frac \partial {\partial x_\mu }\longrightarrow
\gamma _5  \nonumber \\
P_\mu &\longrightarrow &i\frac \partial {\partial x^\mu }\longrightarrow
\gamma _\mu (1-\gamma _5)  \nonumber \\
K_\mu &\longrightarrow &-i(\frac 12x_\nu x^\nu \frac \partial {\partial
x_\mu }-x_\mu x_\nu \frac \partial {\partial x_\nu })\text{ }\longrightarrow
\gamma _\mu (1+\gamma _5)\text{.}  \label{mapping}
\end{eqnarray}
We use $\longrightarrow $ to represent the accurate mappings. The coupling
of the two representations yields the general form
\begin{equation}
\lbrack c_1\gamma _\mu (1-\gamma _5)p^\mu +c_2\frac{\gamma _\mu \gamma _\nu
-\gamma _\nu \gamma _\mu }2M^{\mu \nu }+c_3\gamma _5D+c_4\gamma _\mu
(1+\gamma _5)K^\mu -m_0]\Psi =0,  \label{general}
\end{equation}
and one notes all of these terms become conformal invariant if we
replace the form $A_\mu B^\mu $ with $g_{\mu \nu }A^\mu B^\nu $.
Now we just suppose
the metric tensor $g_{\mu \nu }$ is absorbed temporally by $B^\nu $ as $%
g_{\mu \nu }B^\nu =B_\mu $, so as to complete the following
discussion plainly.

~~\\

We are concerned about a system related to energy loss via
radiation, so we prefer the terms with energy dimension, then in
eq. (\ref{general}) only the first term $\gamma _\mu (1-\gamma
_5)p^\mu $ and the fourth term $\gamma _\mu (1+\gamma _5)K^\mu $
meet this requirement. And we have already employed the operator
$p^\mu $ in Dirac equation in inertial reference frame, so now we
prefer the latter term $\gamma _\mu (1+\gamma _5)K^\mu $. In
addition, we replace $m_0$ with $\triangle m$ in accordance with
the energy loss. It is well known that the operator $K^\mu $
represents the
acceleration of a classical particle in general relativity~\cite{Rohrlich63}%
, but here it represents the differences of momenta between neighboring
points. We choose the operator $K^\mu $ as the form in momentum space, as $%
-i(\frac 12p^2\frac \partial {\partial p_\mu }-p_\mu p_\nu \frac \partial
{\partial p_\nu })$. Now the equation becomes~
\begin{equation}
~\gamma _\mu (1+\gamma _5)K^\mu \,\Psi =\triangle m\,\,\Psi \text{ .}
\label{WaveA}
\end{equation}

~~\\

Physically the energy loss suggests there are no eigen states for operator $%
\gamma _\mu (1+\gamma _5)K^\mu $, which implies $\triangle m=0$ in eq. (\ref
{WaveA}). Then let's check it. Recall that in solving the free Dirac
equation, we first substitute the eigen-function form of momentum operator $%
\not p$, i.e. $u(\mathbf{p})e^{ip\cdot x}$, into $\not p\psi =m_0\psi $,
then we resolve the matrix equation to get the Dirac spinor $u(\mathbf{p})$
related to momentum $\mathbf{p}$. Similarly here we can derive the eigen
function of $\gamma _\mu (1+\gamma _5)K^\mu $. Knowing that $e^{i\frac{%
k\cdot p}{p^2}}$ is the eigen function of $K_\mu =-i(\frac 12p^2\frac
\partial {\partial p_\mu }-p_\mu p_\nu \frac \partial {\partial p_\nu })$,
after substituting $\tilde u(\mathbf{k})\,e^{i\frac{k\cdot p}{p^2}}$($\tilde
u(\mathbf{k})$ is a corresponding spinor) into $\gamma _\mu (1+\gamma
_5)K^\mu \,\Psi =\triangle m\,\Psi $, the remaining part is a matrix
equation $\gamma _\mu (1+\gamma _5)k^\mu \,\tilde u(\mathbf{k})=\triangle
m\,\,\tilde u(\mathbf{k})$, where $k^\mu $'s are eigen values of operators $%
K^\mu $. Writing out the matrix $\gamma _\mu (1+\gamma _5)$ explicitly we
find out $\triangle m\equiv 0$. So somehow there is no eigen function for
matrix operator $\gamma _\mu (1+\gamma _5)k^\mu $, coincident with our
expectation. Then for any spinor $\tilde u(\mathbf{k})$, $\gamma _\mu
(1+\gamma _5)k^\mu \tilde u(\mathbf{k})$ leads to another spinor $\tilde
\upsilon (\mathbf{k})$, as
\begin{equation}
\gamma _\mu (1+\gamma _5)k^\mu \tilde u(\mathbf{k})=k_0\,\tilde \upsilon (%
\mathbf{k})\text{ ,}  \label{Parity A}
\end{equation}
where $k_0$ is a constant accordingly. Multiplying the factor $(1-\gamma _5)$
on both sides of the above equation yields
\[
\gamma _\mu (1+\gamma _5)k^\mu \tilde u_R(\mathbf{k})=k_0\,\tilde \upsilon
_L(\mathbf{k})\text{ ,}
\]
where we define $\tilde u_R(\mathbf{k})=(1+\gamma _5)\tilde u(\mathbf{k})$
and $\tilde \upsilon _L(\mathbf{k})=(1-\gamma _5)\tilde \upsilon (\mathbf{k}%
) $. Or we may write the whole wave functions on both sides as
\begin{equation}
\tilde {\not K}\,\Psi _R(p)=k_0\,\Phi _L(p)\text{ ,}  \label{Parity B}
\end{equation}
where $\tilde {\not K}=\gamma _\mu (1+\gamma _5)K^\mu $, $\Psi _R(p)=\tilde
u_R(\mathbf{k})e^{i\frac{k\cdot p}{p^2}}$, and $\Phi _L(p)=\tilde \upsilon
_L(\mathbf{k})e^{i\frac{k\cdot p}{p^2}}$. And the further inference is
\begin{equation}
\tilde {\not K}\,\Phi _L(p)=0\text{ .}  \label{Parity C}
\end{equation}
In above equations we have employed $(1+\gamma _5)^2=2(1+\gamma
_5)$ and $(1+\gamma _5)\ (1-\gamma _5)=0$. Recalling that Dirac
equation can be separated into two equations like
\begin{eqnarray}
\not p\Psi _R &=&m\,\Psi _L  \nonumber \\
\not p\Psi _L &=&m\,\Psi _R\text{ ,}  \label{DiracAB}
\end{eqnarray}
by which we recognize that the property of eq. (\ref{Parity B}) is totally
different from that of Dirac equation since the existence of eq. (\ref{Parity C}). $%
\tilde {\not K}\,\Phi _L(p)=0$ means somehow a left-handed particle might
break the conformal (scaling) symmetry in strong gravitational field, since
the wave function $\Phi _L(p)$ cannot be acted by any operator with matrix $%
\gamma _\mu (1+\gamma _5)$. $\tilde {\not K}\,\Phi _L(p)=0$ turns out to be
trivial and we need another equation for $\Phi _L(p)$, for a left-handed
fermion moving in gravitational field.

Now the question arises how those of fermions behave which already
evolved to left-handed $\Phi _L(p)$ or, those of neutral particles
behave in strong gravitational field? First, if these particles
take charge, they must also radiate according to the analysis in
introduction part. And without scaling symmetry, such radiation
should depend on which reference frames the observer is sitting
on. Second, as for a micro-neutral particle, where there is still
no co-moving reference frame due to quantum uncertainty, the
particle should also be underlain by an equation breaking
conformal symmetry for its non-radiation. Though the equation
loses the conformal symmetry, i.e. the scaling character, it
certainly still possesses the rotation symmetry which even a free
particle should satisfy. Since any form like $\gamma _\mu Q^\mu
\,\Phi _L(p)=\triangle k\,\Phi _L(p)$ ($Q^\mu $ is an arbitrary
operator) would lead to $\triangle k\,\equiv 0$, the most possible
form is
\begin{equation}
\gamma _\mu K^\mu \,\Phi _L(p)=\gamma _\nu \triangle k^\nu \,\Phi _L(p)
\label{Final A}
\end{equation}
We already know the eigen function of $\gamma _\mu K^\mu $ is $u(\mathbf{k}%
)\,e^{i\frac{k\cdot p}{p^2}}$, where $u(\mathbf{k})$ is the same spinor as
that in free Dirac equation and $\triangle k^\mu $ is the eigen value of $%
K^\mu $, and $\mathbf{k}$ the spatial value of $K^\mu $. Depite of
the spinor $u(\mathbf{k})$, one notes that the only difference
from Dirac equation is the factor $e^{i\frac{k\cdot p}{p^2}}$. We
view the function $u(\mathbf{k} )\,e^{i\frac{k\cdot p}{p^2}}$ as
the eigen function for the above equation however, the unexpected
eigen value form of the equation like $\gamma _\nu \triangle k^\nu
$ surprises us, which may just pertain to particles moving in
gravitational field. While $p\rightarrow \infty $, the drastic
oscillation of $e^{i\frac{k\cdot p}{p^2}}$ leads the wave to a
classical particle.

Below the above equation (\ref{general}) we have mentioned the
metric tensor $g_{\mu \nu }$ is absorbed like $g_{\mu \nu }B^\nu
=B_\mu $. Now we discuss what if writing them explicitly, i.e.
$\gamma _\mu K^\mu \rightarrow g_{\mu \nu }\gamma ^\mu K^\nu $ in
equation (\ref{Final A}), and how to treat them if using the same
set of $\{\gamma _\mu \}$ in difference reference frame.
Geometrically, the metric tensor $g_{\mu \nu }$ is responsible for
the deformation of local bases $\{e_\mu \}$, $g_{\mu \nu }=\langle
e_\mu \,,e_\nu \rangle \,$. The deformation certainly includes the
local inflation and rotations of bases. Equivalently, the term
$g_{\mu \nu }\,K^\nu $ converts the deformation from bases to
vector $\,K^\nu $, just as an additional conformal-transformation
as $\Lambda _{\mu \nu }K^\nu $. Of course such transformation can
also act on $\gamma ^\mu $, as $\gamma ^\mu \Lambda _{\mu \nu }=$
$\gamma _\nu ^{\prime }=S^{-1}\gamma _\nu S$, where $S$ is
responsible for the variation of $\gamma _\mu $ between different
reference frames. And locally, $S$ includes the combination of
rotations, inflations as $c_1e^{\frac u2\gamma _\mu \gamma _\nu
}+c_2e^{\frac u2\gamma _5}$. Apply this transformation to both
sides of eq. (\ref{Final A}), we vary the equation from $\gamma
_\mu K^\mu \,\Phi _L=\gamma _\nu \triangle k^\nu \,\Phi _L$ to
$\gamma _\mu K^\mu \,S\,\Phi _L=\triangle k\,S\,\Phi _L\ $, or
alternatively as
\begin{equation}
(\gamma _\mu K^\mu \,-\gamma _\nu \triangle k^\nu \,)\,\Phi _L^{\prime }=0%
\text{ .}  \label{Final B}
\end{equation}
Finally the effect of space-time curvature is transferred to $\Phi _L$ by $%
\Phi _L^{\prime }=S\,\Phi _L$. So far for different accelerating reference
frames, we need carry additional transformations on the wave function $\Phi
_L$.

Let's give a brief summary for the above construction. First step,
the radiation operator $\gamma _\mu (1+\gamma _5)K^\mu $ (from now
on we name it radiation operator) transforms a right-handed wave
$\Psi _R$ to a left handed $\Phi
_L$. Second step, while a charged left-handed particle (with wave function $%
\Phi _L$) or neutral particle keeps moving, the particle's equation becomes $%
\gamma _\mu K^\mu \,\Phi _L=\gamma _\nu \triangle k^\nu \,\Phi _L$. Third,
the final wave function $\Phi _L$ should experience transformations due to
the variations of metric tensor $g_{\mu \nu }$.

\section{The $CP$ violation caused by the Operator $\tilde {\not K}=\gamma
_\mu (1+\gamma _5)K^\mu $}

Even without resolving the equation eqs. (\ref{Parity B}), (\ref{Parity C})
one can get a critical conclusion. One observes the state in left term is $%
\Psi _R$, and the right term is $\Phi _L$, that means in a
gravitational field strong enough, the acceleration of a charge
would change it helicity. From eq. (\ref{Parity B}) $\tilde {\not
K}\,\Psi _R(p)=k_0\,\Phi _L(p)$ one notes that for a naked wave
function $\Psi $ it gets the effect that a right-handed wave to a
left-handed one. However, for another naked wave purely as
left-handed like $\Phi _L$, then $\tilde {\not K}$ leads $\Phi _L$
to be null according to eq. (\ref{Parity C}). That's means while
we use the operator $\tilde {\not K}$, it makes any wave function
first to be a left-handed, and secondly the left-handed wave
breaks the conformal symmetry but would not be varied by the
operator any longer. While the momentum $p$ is very large, we know
$m/p\rightarrow 0$, then approximately we view the particle
massless. In that case, the projective operator $\frac 12(1\pm
\sigma _p)$ becomes $\frac 12(1\pm \gamma _5)$, the two sets of
operators become equivalent. So, the operator $g_{\mu \nu
}\,\gamma ^\mu K^\nu (1+\gamma _5)$ changing $\Psi _R$ to $\Phi
_L$ is equal to right-handed to left-handed. And for purely
massless particle, to match Dirac picture the left-handed particle
and right-handed particle corresponds to the regular particle and
its anti-particle respectively. Thus the effect of operator
$\tilde {\not K}$ indicates why nowadays normal matter is
dominant. The process might be finished at the first few seconds
of Big Bang of our Universe. And such effect should attribute to
the conformal group, the conformal transformations largely exceed
the scope of the Lorentz transformations, so it could circumvent
the limitation of $ CPT $ theorem to produce $CP$ violation.

\section{A Route to Construct Equation for Operators}

We note the conformal symmetry is mainly manifested by the group
generators as well as operator $K^\mu $. So concerning the
evolution of any operator with quantum paradigm, we need to know
the commutation of this operator with Hamiltonian. While a charge
falling in gravitational field, the Hamiltonian of the charge sets
in an open environment, so it cannot be of a conservative
quantity. It keeps varying under the action of operator $K^\mu $.
Suppose
there is an initial Hamiltonian $\hat H_0$ to be changed under operator $%
K^\mu $, as
\begin{equation}
e^{-i\delta _\mu K^\mu }\hat H_0e^{i\delta _\mu K^\mu }=\hat H_0+i\delta
_\mu [\hat H_0,\hat K^\mu ]\text{ ,}  \label{HamA}
\end{equation}
and from the above commutations in eq. (\ref{Commu}) $\ [P_\mu ,K_\rho
]=-2\,i\,(\eta _{\mu \rho }\,D+M_{\mu \rho })$ one infers that
\[
\lbrack \hat H_0,K_\rho ]=-2\,i\,(\eta _{0\rho }\,D+M_{0\rho })\text{ ,}
\]
or concretely
\begin{eqnarray*}
\lbrack \hat H_0,\hat K_0] &=&-2\,i\,\,D\text{ ,} \\
\lbrack \hat H_0,\hat K_j] &=&-2\,i\,M_{0j}\text{ .}
\end{eqnarray*}
From eq. (\ref{HamA}) the true Hamiltonian after an instant moment is
\begin{eqnarray}
\hat H^{\prime } &=&\hat H_0+i\delta _\mu [\hat H_0,\hat K^\mu ]\text{ }
\nonumber \\
\ &=&\hat H_0+2\,\delta _0\,D+2\,\vec \delta \cdot \vec B\text{ ,}
\label{TrueH}
\end{eqnarray}
where $\vec \delta =(\delta _1,\delta _2,\delta _3)$, and
$B_j=M_{0j}$ means one of the three boosts in corresponding
reference frame, is also a kind of rotation in 4-dimension
space-time. So now for any operator $\hat F$ if we want to know
how it evolves, then just employ the following equation,
\[
i\hbar \frac{d\hat F}{dt}=i\hbar \frac{\partial \hat F}{\partial t}+[\hat
H^{\prime },\hat F]
\]
where $\hat H^{\prime }$ is expressed by eq. (\ref{TrueH}). For example for
the normal angular momentum $\vec L$, it yields
\begin{equation}
i\hbar \frac{d\vec L}{dt}=[\hat H^{\prime },\vec L]\text{ ,}  \label{MagnetA}
\end{equation}
here we assume that $\vec L$ is not implicit varying with time, and
furthermore suppose $[\hat H_0,\vec L]=0$. Then we have
\begin{equation}
i\hbar \frac{d\vec L}{dt}=2[\,\delta _0\,D+\,\vec \delta \cdot \vec B,\,\vec
L]\text{ ,}  \label{MagnetB}
\end{equation}
Using $[\,B_i,\,L_j]=\varepsilon _{ijk}B^k$, the above equation becomes
\[
i\hbar \frac{dL_j}{dt}=2\,\varepsilon _{ijk}\,\delta ^i\,B^k\text{ ,}
\]
i.e.
\begin{equation}
i\hbar \frac{d\vec L}{dt}=2\,\vec B\,\times \vec \delta \text{ .}
\label{MagnetC}
\end{equation}
From this equation the angular momenta keep changing due to the
infinitesimal transformation parameter $\vec \delta $. If only the
Hamiltonian $\hat H$ keeps varying via infinitesimal transformation $%
e^{i\delta _\mu K^\mu }$, then the infinitesimal parameter $\vec
\delta $ attributes to the variation of angular momentum $\vec L$.
This is mimicking very much the movement of perihelion of Mercury
caused by Gravity.

In addition, certainly we can extend the transformation $e^{-i\delta _\mu
K^\mu }\hat H_0e^{i\delta _\mu K^\mu }$ to a complete form maybe like $%
e^{-i\delta \,\gamma _\mu K^\mu }\hat H_0e^{i\delta \,\gamma _\mu K^\mu }$
which surely takes the information of spins. Thus while we carry the
variation of $J=L+S$, $S$ is spin angular momentum, we can find certain spin
part on the right hand side of equation. And we would find $J$ is not a
conservative quantity any longer. Furthermore, one can predict there doesn't
exist a quantity like angular momentum covariant under even conformal
transformations, just like in General Relativity one can not find a
covariant energy-momentum tensor for gravitational field. It is deep-rooted
in the conformal transformations.

~~\\

\section{Conclusions and Discussions}

In this paper we have carried out a quantum equation for falling
fermion in strong gravitational field from the initial hypothesis
of conformal symmetry. And we observe that the operator $\tilde
{\not K}$ causes the conformal symmetry broken spontaneously.
Subsequently the breaking of conformal symmetry makes the EoM of
charged fermions varied to a familiar one like Dirac equation,
with operator $K$ in the place of $p$. The unexpected aspect is
the tensor-like eigen value. We have also discussed how the metric
tensors $g_{\mu \nu }$ affect the wave function. We have obtained
the surprising conclusion that the final observation of radiation
still depends on reference frame, partially in agreement with
other authors who discussed the issue from a classical viewpoint.
From the property of operator $\tilde {\not K}$ we find the
possible reason for strong CP violation at the very beginning of
Big Bang.

~~\\

With this paper's progress, one may still stick to the perception
of a comoving reference frame or an identical accelerated
reference frame to let a charge move freely. For example put a net
uniform electronic field in a curved space-time, according to the
analysis in ref.~\cite{Soker99}, the field would radiate. If let
the field fall freely, then it seems certain the radiation would
disappear. Here, the loophole of the thought is the neglect of the
fact that the field lines are spatially extended, nonlocal~\cite
{Rohr08}. So it is impossible to arrange a comoving
infinitesimally local reference frame for all of the field lines
as a whole, i.e. the local flat-space-time condition required by
GR cannot be found by shrinking the scale of the spacetime. This
point has been implied by quite a few recent researches
~\cite{Crem15,Khok18,Singal17,Harte12,Harte18}. To speak
alternatively, the electronic field is spreading and thus a
nonlocal entity. The analog exists around electron, as well as all
of fermions.

~~\\

One might also perceive that the radiation energy is just the work
gravity acting on the mass of the charge, like acting on a
classical particle, which is easy to carry out. However, with
above arguments relating to nonlocality, the calculation becomes
almost impossible since the mass (energy) should also include
those of electromagnetic field lines. It touches the hard shell of
the question, i.e. how gravity interacts with an electromagnetic
entity. One does know how the gravity interacts with mass part of
the electromagnetic entity, but one doesn't know how gravity
interacts with the charge or the field lines. Without the
interaction details, we may still be able to evaluate the
radiation form by carrying out the difference of wave functions of
two neighboring points in curved spacetime. We can evaluate the
probability of charge radiation, which according to the wave
functions is in proportion to radiation power $\mid
e^{i\frac{k\cdot p^{\prime }}{p^{\prime 2}}}-e^{i\frac{k\cdot
p}{p^2}}\mid ^2\propto k^2\propto a^2$, where $a$ is classical
acceleration. And the well known radiation power is actually in
proportion to $a^2$~\cite{Jackson}. The further comparison is in
progress. If this method is available, we can also extract the
information of the gravitational field of remote stars by the
electromagnetic radiation around them~\cite{experiment19}.

~~\\

If only a charged particle is accelerated, it would radiate due to
bremsstrahlung. So we may design experiment at cyclotron to test
our predictions in this paper. The predictions include the change
of right to left-handed, the variations of angular momentum, and
the reference-frame-dependence of the radiation, etc.. Before
designing the experiment, first of all we should differentiate the
two radiations of a charge caused by electromagnetic field and by
gravitational field (accelerations) respectively. While a charge
being accelerated by electric field, it interacts with the
electric field, thus part of radiation is directly caused by the
interaction. The interaction causes some kinds of transition
between quantum states, so the radiation energies are some
regularly distributions submitted to Boltzmann law. Here we define
such radiation as background. At the same time, while a charge
being accelerated (in any manner), the field lines dressed by the
charge would definitely be deformed, then the bremsstrahlung
happens due to the induction-like effect, which has a continuous
spectrum in proportion to $a^2$. These two radiation mechanisms
are totally different. Therefore, to find the true bremsstrahlung
radiation related to acceleration, we should extract the radiation
from the conventional bremsstrahlung by getting rid of the
background part, i.e. Boltzmann part.

~~\\

\begin{acknowledgments}
The work is supported in part by National Science Foundation of
China (NSFC) under Grant No. 11647304, 11475071, and 11547308.
\end{acknowledgments}

~~\\

\end{document}